\documentstyle[12pt]{article}

\def\thebibliography#1{\section*{References}\list
  {[\arabic{enumi}]}{\settowidth\labelwidth{#1}\leftmargin\labelwidth
    \advance\leftmargin\labelsep
    \usecounter{enumi}}
    \def\newblock{\hskip .11em plus .33em minus .07em}
    \sloppy\clubpenalty4000\widowpenalty4000
    \sfcode`\.=1000\relax}

\def\op#1{\mathop{\fam0 #1}\limits}

\newcommand{\nm}[1]{\mid {#1}\mid}

\newcommand{\beq}{\begin{equation}}
\newcommand{\eeq}{\end{equation}}
\newcommand{\ben}{\begin{eqnarray}}
\newcommand{\een}{\end{eqnarray}}
\newcommand{\be}{\begin{eqnarray*}}
\newcommand{\ee}{\end{eqnarray*}}
\newcommand{\bea}{\begin{eqalph}}
\newcommand{\eea}{\end{eqalph}}

\newcommand{\cA}{{\cal A}}

\newcommand{\cV}{{\cal V}}

\newcommand{\cL}{{\cal L}}
\newcommand{\cE}{{\cal E}}
\newcommand{\cH}{{\cal H}}

\newcommand{\bL}{{\bf L}}
\newcommand{\bR}{{\bf R}}
\newcommand{\bC}{{\bf C}}

\newcommand{\bt}{\beta}
\newcommand{\dl}{\delta}
\newcommand{\la}{\lambda}
\newcommand{\La}{\Lambda}
\newcommand{\f}{\phi}
\newcommand{\om}{\omega}
\newcommand{\Om}{\Omega}
\newcommand{\m}{\mu}

\newcommand{\g}{\gamma}
\newcommand{\G}{\Gamma}

\newcommand{\si}{\sigma}
\newcommand{\w}{\wedge}
\newcommand{\wt}{\widetilde}
\newcommand{\wh}{\widehat}
\newcommand{\ol}{\overline}
\newcommand{\dr}{\partial}

\newcommand{\ot}{\otimes}

\newcommand{\nw}[1]{[{#1}]}
\newcommand{\der}{\rm Der}
\newcommand{\pr}{{\rm pr}}

\newcounter{eqalph}
\newcounter{equationa}

\newenvironment{eqalph}{\stepcounter{equation}
\setcounter{equationa}{\value{equation}}
\setcounter{equation}{0}

\begin{eqnarray}}{\end{eqnarray}\setcounter{equation}{\value{equationa}}}

\begin{document}
\hbox{}

\begin{center}
{\large\bf  SUSY-EXTENDED FIELD THEORY}
\bigskip

{\sc GENNADI SARDANASHVILY}\footnote{E-mail: sard@grav.phys.msu.su}

{\small
 Department of Theoretical Physics, Physics Faculty, Moscow State
University,

117234 Moscow, Russia}
\bigskip

\end{center}

{\small
A field model on fibre bundles can be extended in a
standard way
to the BRS-invariant SUSY field model which possesses the Lie
supergroup ISp(2) of symmetries.  }

\section{Introduction}

Generalizing the BRS
mechanics of E.Gozzi and M.Reuter [1-4], we show that any 
field model (independently of its physical
symmetries) can be extended 
in a standard way to 
a SUSY field model. In comparison with the SUSY field
theory in Ref. [5,6], this extension is formulated in terms of simple graded
manifolds. From the physical viewpoint, the SUSY-extended field theory
may describe odd deviations of physical fields, e.g., of a Higgs field.

We follow the conventional geometric formulation of field
theory  where classical fields are represented by
sections of a smooth fibre bundle $Y\to X$ coordinated by $(x^\la,y^i)$. 
 A first order Lagrangian $L$ of field theory is 
defined as a horizontal  density
\beq
L=\cL(x^\la,y^i,y^i_\la)\om: J^1Y\to\op\w^nT^*X, \quad \om=dx^1\w\cdots
\w dx^n, \quad n=\dim X, 
\label{cmp1}
\eeq
on the first order jet manifold $J^1Y\to Y$ of sections of $Y\to X$
[7-9]. The jet 
manifold $J^1Y$ coordinated by $(x^\la,y^i,y^i_\la)$ plays the role of a
finite-dimensional configuration 
space of fields.
The corresponding
Euler--Lagrange equations take the coordinate form
\beq
\dl_i\cL=(\dr_i- d_\la\dr^\la_i)\cL=0, \qquad d_\la=\dr_\la +y^i_\la\dr_i
+y^i_{\la\m}\dr^\m_i.
\label{b327}
\eeq
Every Lagrangian $L$ (\ref{cmp1}) yields the Legendre map 
\be
\wh L: J^1Y\op\to_Y \Pi, \qquad p^\la_i\circ\wh L=\pi_i^\la=\dr^\la_i\cL,
\ee
 of
$J^1Y$ to the Legendre bundle 
\beq
\Pi=\op\w^nT^*X\op\ot_YV^*Y\op\ot_YTX, \label{00}
\eeq
where $V^*Y$ is the vertical cotangent bundle of $Y\to X$.
 The Legendre bundle $\Pi$ is equipped with the holonomic coordinates
$(x^\la,y^i,p^\la_i)$. It is
seen as a momentum
phase space of covariant Hamiltonian field theory where canonical
momenta correspond to derivatives of fields with respect to all
space-time coordinates [8-10]. 
Hamiltonian dynamics on $\Pi$ is phrased in terms of Hamiltonian forms
\beq
 H= p^\la_i dy^i\w \om_\la -\cH(x^\la,y^i,p^\la_i)\om, \qquad
\om_\la=\dr_\la\rfloor\om. \label{b418}
\eeq
The corresponding covariant Hamilton equations read
\beq
y^i_\la= \dr_\la^i\cH, \qquad p^\la_{\la i}= - \dr_i\cH. \label{b4100}
\eeq

Note that, if $X={\bf R}$, covariant Hamiltonian formalism provides the
adequate Hamiltonian formulation of time-dependent mechanics
\cite{book98,sard98}. This fact enables us 
to extend the above-mentioned BRS mechanics of E.Gozzi and M.Reuter to
field theory.

A preliminary step toward the desired SUSY extension of field theory is
its extension to field theory on the vertical tangent bundle
$VY$ of $Y\to X$, coordinated by $(x^\la,y^i,\dot y^i)$ \cite{book}. 
Coordinates 
$\dot y^i$ describe linear deviations of fields. The configuration and
phase spaces of field theory on $VY$ are 
\ben
&& J^1VY\cong VJ^1Y, \qquad (x^\la,y^i, y^i_\la,\dot y^i, \dot y^i_\la),
\label{ijmp1}\\ 
&& \Pi_{VY}\cong V\Pi, \qquad (x^\la,y^i,p_i^\la,\dot y^i, \dot
p_i^\la).\label{ijmp2} 
\een
Due to the isomorphisms (\ref{ijmp1}) -- (\ref{ijmp2}), the
corresponding vertical prolongation of the Lagrangian $L$ (\ref{cmp1})
reads 
\beq
L_V=(\dot y^i\dr_i + \dot y^i_\la\dr_i^\la)L, \label{ijmp3}
\eeq
while that of the Hamiltonian form $H$ (\ref{b418}) is
\beq
H_V=(\dot p^\la_idy^i + p^\la_i d\dot
y^i)\w\om_\la -\cH_V\om, \qquad \cH_V =(\dot y^i\dr_i +\dot
p^\la_i\dr^i_\la)\cH.
\label{m17}
\eeq
The corresponding Euler--Lagrange and Hamilton equations describe the
Jacobi fields of solutions of the Euler--Lagrange equations
(\ref{b327}) and the Hamilton equations (\ref{b4100}) of the initial
field model on $Y$. 

The SUSY-extended field theory is formulated in terms of simple graded
manifolds whose characteristic vector bundles are the vector bundles
$VJ^1VY\to J^1VY$ in Lagrangian formalism and  $V\Pi_{VY}\to
\Pi_{VY}$ in Hamiltonian formalism \cite{nuo}. The SUSY extension adds to
$(y^i,
\dot y^i)$
the odd variables $(c^i,\ol c^i)$. The corresponding SUSY extensions of 
the Lagrangian $L$ (\ref{cmp1}) and the Hamiltonian form $H$
(\ref{b418}) are constructed in order to be invariant under the 
BRS transformation
\beq
u_Q= c^i\dr_i + i\dot y^i
\frac{\dr}{\dr \ol c^i}. \label{ijmp30}
\eeq
Moreover, let us consider fibre bundles $Y\to X$ characterized by
affine transition functions of bundle coordinates $y^i$ (they are not
necessarily affine bundles). 
Almost all  
field models are of this type.  
In this case,
the transition functions of holonomic coordinates
$\dot y^i$ on $VY$ are independent of $y^i$, and 
$(c^i)$ and $(\ol c^i)$ have the same transition functions. 
Therefore, BRS-invariant Lagrangians and Hamiltonian forms of the
SUSY-extended field theory are also invariant under the 
Lie supergroup ISp(2) with the generators $u_Q$ (\ref{ijmp30}) and
\beq
u_{\ol Q}=\ol c^i\dr_i  -i\dot y^i
\frac{\dr}{\dr c^i},  \quad
u_K=c^i\frac{\dr}{\dr \ol c^i}, \quad
u_{\ol K}=\ol c^i\frac{\dr}{\dr c^i},
\quad u_C=c^i\frac{\dr}{\dr c^i}  -\ol c^i\frac{\dr}{\dr\ol c^i}.
\label{cmp131}
\eeq

Note that, since the
SUSY-extended field theory, by construction, is BRS-invariant, one
hopes that its quantization may be free from some divergences.

\section{Technical preliminaries}

Given a fibre bundle $Z\to X$ coordinated by $(x^\la,z^i)$, the 
$k$-order jet manifold $J^kZ$ is endowed with the adapted 
coordinates $(x^\la,z^i_\La)$, $0\leq\mid\La\mid\leq k$, where 
$\La$ is a symmetric multi-index $(\la_m...\la_1)$, $\nm\La=m$. 
Recall the canonical morphism
\be
\la=dx^\la \otimes (\dr_\la + z^i_\la \dr_i):J^1Z\op\hookrightarrow_Y T^*X
\op\ot_Z TZ. 
\ee

Exterior forms $\f$ on a jet manifold $J^kZ$, $k=0,1,\ldots$, are naturally
identified with their pull-backs onto $J^{k+1}Z$. There is the exterior algebra
homomorphism
\beq
h_0:\f_\la dx^\la+ \f_i dz^i \mapsto (\f_\la+\f_i^\La y^i_\la)dx^\la,
\label{cmp100} 
\eeq
 called the horizontal projection, which sends exterior forms on $J^kZ$
onto the horizontal forms on 
$J^{k+1}Z\to X$. Recall the operator of 
the total derivative 
\be
d_\la =\dr_\la +z^i_\la\dr_i + z^i_{\la\m}\dr_i^\m
+\cdots.
\ee

A connection on a fibre bundle $Z\to X$ is regarded as a
global section 
\be
\G=dx^\la\ot(\dr_\la +\G^i_\la\dr_i) 
\ee
of the affine jet bundle $J^1Z\to Z$. One says that a section
$s:X\to Z$ is an integral section of $\G$ if
$\G\circ s=J^1s$, where $J^1s$ is the jet prolongation of $s$ to a
section of the jet bundle $J^1Z\to X$.

Given the vertical tangent bundle $VZ$ of a fibre bundle
$Z\to X$, we will use the notation
\be
\dot\dr_i=\frac{\dr}{\dr \dot  z^i}, \qquad \dr_V= \dot z^i\dr_i.
\ee

\section{The covariant Hamiltonian field theory}

Let us summarize briefly the basics of the covariant Hamiltonian field theory 
on the Legendre bundle $\Pi$ (\ref{00}) [8-10].

Let $Y\to X$ be a fibre bundle and $L$ a Lagrangian (\ref{cmp1}) on $J^1Y$.
The associated Poincar\'e--Cartan form 
\be
H_L=\pi^\la_i dy^i\w\om_\la -(\pi^\la_i y^i_\la -\cL)\om
\ee 
is defined as the horizontal Lepagean equivalent
of $L$ on $J^1Y\to Y$, i.e. $L=h_0(H_L)$, where $h_0$ is the horizontal
projection (\ref{cmp100}).  
Every
Poincar\'e--Cartan form
$H_L$ yields the bundle morphism of
$J^1Y$ to the homogeneous Legendre bundle
\beq
Z_Y=J^{1\star}Y = T^*Y\w(\op\w^{n-1}T^*X)  \label{N41}
\eeq
equipped with the holonomic coordinates
$(x^\la,y^i,p^\la_i,p)$. 
Because of the canonical isomorphism $\Pi\cong
V^*Y\op\w_Y(\op\w^{n-1}T^*X)$, we have the 1-dimensional affine bundle
\beq
\pi_{Z\Pi}:Z_Y\to \Pi. \label{b418'}
\eeq

The homogeneous Legendre bundle
$Z_Y$ is provided with the canonical
exterior $n$-form
\beq
\Xi_Y= p\om +p^\la_idy^i\om_\la. \label{N43}
\eeq
Then a Hamiltonian form $H$ (\ref{b418}) on the Legendre bundle $\Pi$
is defined in an intrinsic way as the pull-back $H=h^*\Xi_Y$ of this
canonical form by some section
$h$ of the fibre bundle (\ref{b418'}). It is readily observed that the
Hamiltonian form $H$ 
(\ref{b418}) is the Poincar\'e--Cartan form of the 
Lagrangian
\beq
L_H=h_0(H) = (p^\la_iy^i_\la - \cH)\om \label{Q3}
\eeq
on the jet manifold $J^1\Pi$, and the Hamilton equations (\ref{b418})
coincide with the Euler--Lagrange equations for this Lagrangian. Note
that the
Lagrangian $L_H$  plays a prominent role in the path integral
formulation of Hamiltonian mechanics and field theory.

The Legendre bundle $\Pi$ (\ref{00}) is provided with the canonical
polysymplectic form
\beq
\Om_Y =dp_i^\la\w dy^i\w \om\ot\dr_\la. \label{406}
\eeq
Given a Hamiltonian form $H$ (\ref{b418}), a connection 
\beq
\g =dx^\la\otimes(\dr_\la +\g^i_\la\dr_i
+\g^\m_{\la i}\dr^i_\m) \label{cmp33}
\eeq
on $\Pi\to X$ is called a Hamiltonian
connection for $H$  if it obeys the condition
\ben
&& \g\rfloor\Om_Y= dH, \nonumber\\
&& \g^i_\la =\dr^i_\la\cH, \qquad 
\g^\la_{\la i}=-\dr_i\cH. \label{3.10}
\een
Every integral section $r:X\to\Pi$ of the Hamiltonian connection $\g$
(\ref{3.10}) is a solution of the Hamilton equations (\ref{b4100}).
Any Hamiltonian form $H$ admits a Hamiltonian connection $\g_H$
(\ref{3.10}). A Hamiltonian form $H$ (\ref{b418}) defines the
Hamiltonian map 
\be
\wh H: \Pi \op\to^{\g_H} J^1\Pi \op\to_Y J^1Y, \qquad y^i_\la=\dr_\la^i\cH,
\ee
which is the same for all Hamiltonian connections associated with $H$.

In the case of hyperregular Lagrangians, Lagrangian and
covariant Hamiltonian formalisms are equivalent. For any hyperregular
Lagrangian $L$, there exists a unique Hamiltonian form $H$ such that the
Euler--Lagrange equations (\ref{b327}) for $L$ are equivalent to the
Hamilton equations (\ref{b4100}) for $H$. The case of degenerate Lagrangians
is more intricate. One can state certain relations between solutions of
Euler--Lagrange and Hamilton equations if a Hamiltonian form $H$ 
satisfies the relations
\bea
&& \wh L\circ \wh H\circ\wh L=\wh L, \label{2.30a}\\
&& p^\la_i\dr_\la^i\cH-\cH=\cL\circ\wh H \label{2.30b}
\eea
[8-10]. It is called associated with a Lagrangian $L$.
If the relation (\ref{2.30b}) takes place on the Lagrangian
constraint space $\wh L(J^1Y)\subset \Pi$, one says that $H$ 
is weakly associated with $L$.  We will show that,
if $L$ and $H$ are associated, then their vertical and SUSY
prolongations are weakly associated. 

\section{The vertical extension of field theory}

As was mentioned above, the vertical extension of field theory on a
fibre bundle $Y\to X$ to 
the vertical tangent bundle $VY\to X$ describes linear deviations of
fields. 

Let $L$ be a Lagrangian on the configuration space $J^1Y$. Due to the
isomorphism (\ref{ijmp1}), its prolongation (\ref{ijmp3}) onto the vertical
configuration space $J^1VY$ can be defined as the vertical tangent morphism
\ben
&& L_V=\pr_2\circ VL: VJ^1Y\to \op\w^nT^*X, \label{m18}\\
&& \cL_V= \dr_V\cL=(\dot y^i\dr_i + \dot y^i_\la\dr_i^\la)\cL, \nonumber
\een
to the morphism $L$ (\ref{cmp1}). The corresponding Euler--Lagrange
equations read
\bea
&& \dot\dl_i \cL_V=\dl_i\cL=0, \label{ijmp6}\\
&& \dl_i\cL_V=\dr_V\dl_i\cL =0, \qquad \dr_V=\dot y^i\dr_i + \dot
y^i_\la \dr_i^\la + \dot y^i_{\m\la}\dr_i^{\m\la}. \label{ijmp7}
\eea
The equations (\ref{ijmp6}) are exactly the
Euler--Lagrange equations (\ref{b327}) for the Lagrangian $L$. In order
to clarify the physical meaning of the equations (\ref{ijmp7}), let us
suppose that $Y\to X$ is a vector bundle. Given a solution $s$ of the
Euler--Lagrange equations (\ref{b327}), let $\dl s$ be a Jacobi field,
i.e., $s+\varepsilon\dl s$ is also a solution of the same Euler--Lagrange
equations modulo terms of order $>1$ in a parameter $\varepsilon$. Then
it is readily 
observed that the Jacobi field $\dl s$ satisfies the Euler--Lagrange
equations (\ref{ijmp7}). 

The Lagrangian (\ref{m18}) yields the vertical Legendre map
\ben
&& \wh L_V=V\wh L: VJ^1Y\op\to_{VY} V\Pi, \label{ijmp56}\\
&& p^\la_i=\dot\dr^\la_i\cL_V=\pi^\la_i, \qquad
\dot p^\la_i =\dr_V\pi^\la_i. \label{ijmp57}
\een
Due to the isomorphism (\ref{ijmp2}), the vertical tangent bundle
$V\Pi$ of $\Pi\to X$ plays the role of the momentum phase space of
field theory on $VY$, where
the canonical conjugate pairs are
$(y^i,\dot p^\la_i)$ and $(\dot y^i,p_i^\la)$. In particular, 
$V\Pi$ is endowed with the canonical polysymplectic form 
(\ref{406}) which reads
\beq
\Om_{VY}=\dr_V\Om_Y=[d\dot p^\la_i\w dy^i +dp^\la_i\w d\dot
y^i]\w\om\ot\dr_\la. 
\label{cmp35}
\eeq

Let $Z_{VY}$ be the homogeneous Legendre bundle (\ref{N41}) over $VY$ 
with the corresponding coordinates $(x^\la,y^i,\dot
y^i,p_i^\la,q_i^\la,p)$. It can be endowed with  
the canonical form $\Xi_{VY}$ (\ref{N43}). Sections of the
affine bundle
\beq
 Z_{VY}\to V\Pi, \label{cmp41}
\eeq
by definition, provide Hamiltonian forms on $V\Pi$. Let us
consider the following particular case of these forms which are
related to those on the Legendre bundle $\Pi$.
Due to the fibre bundle
\ben
&&\zeta: VZ_Y\to Z_{VY}, \label{cmp40}\\
&& (x^\la,y^i,\dot y^i,p_i^\la,q_i^\la,p) \circ\zeta= (x^\la,y^i,\dot
y^i,\dot p_i^\la, p_i^\la,\dot p), \nonumber
\een
the vertical tangent bundle $VZ_Y$ of $Z_Y\to X$ is provided with
the exterior form 
\be
\Xi_V=\zeta^*\Xi_{VY}= \dot p\om + (\dot p^\la_i
dy^i + p^\la_id\dot y^i) \w\om_\la.
\ee
Given the affine bundle 
$Z_Y\to\Pi$ (\ref{b418'}), we have the fibre bundle
\beq
V\pi_{Z\Pi}: VZ_Y\to V\Pi, \label{cmp34}
\eeq
where $V\pi_{Z\Pi}$ is the vertical tangent map to $\pi_{Z\Pi}$. The fibre
bundles (\ref{cmp41}), (\ref{cmp40}) and (\ref{cmp34}) form the commutative
diagram.
Let $h$ be a section of the affine bundle $Z_Y\to \Pi$ and
$H=h^*\Xi$ the corresponding Hamiltonian form (\ref{b418}) on $\Pi$.  
Then the section $Vh$ of the fibre bundle (\ref{cmp34}) and the
corresponding section $\zeta\circ Vh$ of the affine bundle (\ref{cmp41})
defines the Hamiltonian form $H_V=(Vh)^*\Xi_V$ (\ref{m17}) on $V\Pi$. One can
think of this form as being a vertical extension of $H$. In particular,
let $\g$ (\ref{cmp33}) be a Hamiltonian connection on $\Pi$
for the Hamiltonian form $H$.
Then its vertical prolongation 
\be
V\g=\g + dx^\m\ot [\dr_V\g^i_\m\dot\dr_i +\dr_V\g^\la_{\m i}\dot\dr_\la^i].
\ee
on $V\Pi\to X$ is a Hamiltonian connection for the 
vertical Hamiltonian form $H_V$ (\ref{m17}) with respect to the
polysymplectic form $\Om_{VY}$ (\ref{cmp35}).

The Hamiltonian form
$H_V$ (\ref{m17}) defines the 
Lagrangian $L_{H_V}$ (\ref{Q3}) on $J^1V\Pi$, which takes the form
\be
L_{H_V}=h_0(H_V)=[\dot p^\la_i(y^i_\la- \dr^i_\la\cH) -\dot y^i(p^\la_{\la
i} + \dr_i\cH) +d_\la(p^\la_i\dot y^i)]\om. 
\ee
The corresponding Hamilton equations contain the Hamilton equations
(\ref{b4100}) and the equations
\be
\dot y^i_\la=\dr^i_\la\cH_V=\dr_V\dr^i_\la\cH,\qquad
 \dot p^\la_{\la i}=-\dr_i\cH_V=-\dr_V\dr_i\cH 
\ee
for Jacobi fields $\dl y^i=\varepsilon \dot y^i$, $\dl p^\la_i=
\varepsilon \dot p^\la_i$.

The Hamiltonian form $H_V$ 
(\ref{m17}) on $V\Pi$ yields the vertical
Hamiltonian map
\ben
&& \wh H_V=V\wh H: V\Pi\op\to_{VY} VJ^1Y, \label{ijmp58}\\
&& y^i_\la=\dot\dr^i_\la\cH_V =\dr^i_\la\cH, \qquad
\dot y^i_\la= \dr_V\dr^i_\la\cH. \label{ijmp59}
\een
Let $H$ be associated with a Lagrangian $L$. Then $H_V$ is weakly
associated with $L_V$. Indeed, if the morphisms $\wh H$ and $\wh L$
obey the relation (\ref{2.30a}), then 
the corresponding vertical tangent morphisms satisfy the relation
\be
V\wh L\circ V\wh H\circ V\wh L=V\wh L.
\ee
The condition (\ref{2.30b}) for $H_V$ reduces to the equality
\be
\dr_i\cH(p)=-(\dr_i\cL\circ\wh H)(p), \qquad p\in \wh L(J^1Y),
\ee
which is
fulfilled if $H$ is associated with $L$ [8-10].

\section{Geometry of simple graded manifolds}

Following the BRS extension of Hamiltonian mechanics, we formulate the
SUSY-extended field theory in terms of simple graded manifolds.

Let 
$E\to Z$ be a vector bundle with an $m$-dimensional
typical fibre $V$ and $E^*\to Z$ the dual of $E$. Let us
consider the exterior bundle
\beq
\w E^*=\bR\op\oplus_Z(\op\oplus_{k=1}^m\op\w^k E^*) \label{z780}
\eeq
whose typical fibre is the finite Grassmann algebra $\w V^*$. By $\cA_E$
is meant the sheaf of its sections. The pair $(Z,\cA_E)$ is a
graded manifold with the body manifold $Z$ coordinated by $(z^A)$ and
the structure sheaf 
$\cA_E$ [13-15]. We agree to call it a simple graded manifold with the
characteristic vector bundle $E$. 
This is not the terminology of \cite{cari97} where this term is applied
to all graded manifolds of finite rank in connection with Batchelor's
theorem. In accordance with this theorem, any graded manifold is
isomorphic to a simple graded manifold though this isomorphism is not
canonical \cite{bart,batch1}. To keep the structure of a simple graded
manifold, we will restrict transformations of $(Z,\cA_E)$ to those induced by
bundle automorphisms of $E\to Z$.

Global sections of the exterior bundle (\ref{z780}) are called
graded functions. They make up the
${\bf Z}_2$-graded ring $\cA_E(Z)$.  
Let $\{c^a\}$ be a basis for $E^*\to Z$ with respect to some bundle
atlas with transition functions $\{\rho^a_b\}$, i.e.,
$c'^a=\rho^a_b(z)c^b$. We will call $(z^A, c^i)$ the local basis for the
simple graded manifold $(Z,\cA_E)$. With respect to this basis, graded
functions read 
\beq
f=\op\sum_{k=0}^m \frac1{k!}f_{a_1\ldots
a_k}c^{a_1}\cdots c^{a_k}, \label{z785}
\eeq
where $f_{a_1\cdots
a_k}$ are local functions on $Z$, and we omit the symbol of the
exterior product 
of elements $c$. The coordinate transformation law of graded functions
(\ref{z785}) is obvious. We will use the
notation $\nw .$ of the Grassmann parity.

Given a simple graded manifold $(Z,\cA_E)$, by the sheaf $\der\cA_E$ 
of graded derivations
of $\cA_E$ is meant a subsheaf of endomorphisms of $\cA_E$ such that any 
its section
$u$ over an open subset $U\subset Z$ is a graded derivation of the
graded algebra $\cA_E(U)$ of local sections of the exterior bundle $\w
E^*|_U$, i.e.,  
\be
 u(ff')=u(f)f'+(-1)^{\nw u\nw f}fu (f') 
\ee
for the homogeneous elements $u\in(\der\cA_E)(U)$ and $f,f'\in \cA_E(U)$.
Graded derivations are called graded vector fields on a graded
manifold $(Z,\cA_E)$ (or simply on Z if there is no danger of
confusion). They 
can be represented by sections of some vector bundle as follows.

Due to the canonical splitting
$VE\cong E\times E$, the vertical tangent bundle 
$VE\to E$ can be provided with the fibre basis $\{\dr/\dr c^a\}$ dual of
$\{c^a\}$. 
This is the fibre basis for $\pr_2VE\cong E$.  Then
a graded vector field on a trivialization domain $U$ reads
\be
u= u^A\dr_A + u^a\frac{\dr}{\dr c^a},
\ee
where $u^A, u^a$ are local graded functions.
It yields a derivation of $\cA_E(U)$ by the rule
\beq
u(f_{a\ldots b}c^a\cdots c^b)=u^A\dr_A(f_{a\ldots b})c^a\cdots c^b +u^d
f_{a\ldots b}\frac{\dr}{\dr c^d}\rfloor (c^a\cdots c^b). \label{cmp50'}
\eeq
This rule implies the corresponding
coordinate transformation law 
\be
u'^A =u^A, \qquad u'^a=\rho^a_ju^j +u^A\dr_A(\rho^a_j)c^j 
\ee
of graded vector fields. It follows that graded vector fields on $Z$
can be represented by
sections of the vector bundle
$\cV_E\to Z$ which is locally isomorphic to the vector bundle
\be
\cV_E\mid_U\approx\w E^*\op\ot_Z(\pr_2VE\op\oplus_Z TZ)\mid_U,
\ee
and has the transition functions
\be
&& z'^A_{i_1\ldots i_k}=\rho^{-1}{}_{i_1}^{a_1}\cdots
\rho^{-1}{}_{i_k}^{a_k} z^A_{a_1\ldots a_k}, \\
&& v'^i_{j_1\ldots j_k}=\rho^{-1}{}_{j_1}^{b_1}\cdots
\rho^{-1}{}_{j_k}^{b_k}\left[\rho^i_jv^j_{b_1\ldots b_k}+ \frac{k!}{(k-1)!} 
z^A_{b_1\ldots b_{k-1}}\dr_A\rho^i_{b_k}\right] 
\ee
of the bundle coordinates $(z^A_{a_1\ldots a_k},v^i_{b_1\ldots b_k})$,
$k=0,\ldots,m$. These transition functions
fulfill the cocycle relations. 
Graded vector fields on $Z$ constitute a Lie
superalgebra with respect to the bracket 
\be
[u,u']=uu' + (-1)^{\nw u\nw{u'}+1}u'u.
\ee

There is the exact sequence over $Z$ of vector
bundles
\be
0\to \w E^*\op\ot_Z\pr_2VE\to\cV_E\to \w E^*\op\ot_Z TZ\to 0. 
\ee
Its splitting 
\beq
\wt\g:\dot z^A\dr_A \mapsto \dot z^A(\dr_A +\wt\g_A^a\frac{\dr}{\dr
c^a}) \label{cmp70} 
\eeq
is represented by a section
\beq
\wt \g=dz^A\ot(\dr_A +\wt\g^a_A\frac{\dr}{\dr c^a}) \label{cmp93}
\eeq
of the vector bundle $T^*Z\op\ot_Z\cV_E\to Z$ such that the composition
\be
Z\op\to^{\wt\g}T^*Z\op\ot_Z\cV_E\to T^*Z\op\ot_Z (\w
E^*\op\ot_Z TZ)\to T^*Z\op\ot_ZTZ
\ee
is the canonical form $dz^A\ot\dr_A$ on $Z$.
The splitting (\ref{cmp70}) transforms every vector field $\tau$ on $Z$
into a graded vector field 
\beq
\tau=\tau^A\dr_A\mapsto \nabla_\tau=\tau^A(\dr_A
+\wt\g_A^a\frac{\dr}{\dr c^a}), 
\label{ijmp10} 
\eeq
which is a graded derivation of $\cA_E$ satisfying the Leibniz rule
\be
\nabla_\tau(sf)=(\tau\rfloor ds)f +s\nabla_\tau(f), \quad f\in\cA_E(U),
\quad s\in C^\infty(Z),\quad \forall U\subset Z.
\ee
Therefore, one can think of the graded derivation $\nabla_\tau$
(\ref{ijmp10}) and, consequently, of the splitting (\ref{cmp70}) as
being a graded connection on the simple graded manifold $(Z,\cA_E)$.
In particular, this connection provides the corresponding decomposition
\be  
u= u^A\dr_A + u^a\frac{\dr}{\dr c^a}=u^A(\dr_A +\wt\g_A^a\frac{\dr}{\dr
c^a}) + (u^a- 
u^A\wt\g_A^a)\frac{\dr}{\dr c^a}
\ee
of graded vector fields on $Z$. 
Note that this notion of a graded connection differs from that of
connections on graded fibre bundles in Ref. \cite{lop}. 

In accordance
with the well-known theorem on a splitting of an exact sequence of vector
bundles, graded connections always exist.
For instance, every linear connection 
\be
\g=dz^A\ot (\dr_A +\g_A{}^a{}_bv^b\dr_a) 
\ee
on the vector bundle $E\to Z$ yields the graded connection 
\beq
\g_S=dz^A\ot (\dr_A +\g_A{}^a{}_bc^b\frac{\dr}{\dr c^a}) \label{cmp73}
\eeq
such that, for any vector field $\tau$ on $Z$ and any graded function $f$,
the graded derivation $\nabla_\tau(f)$ with respect to the connection
(\ref{cmp73}) is exactly the covariant derivative
of $f$ relative to the connection $\g$.

Let now $Z\to X$ be a fibre bundle coordinated by $(x^\la,z^i)$,  and let
\be
\g=\G +\g_\la{}^a{}_bv^bdx^\la\ot\dr_a
\ee
be a connection on $E\to X$ which is a linear morphism over a connection $\G$
on $Z\to X$. Then we have the bundle monomorphism
\be
\g_S: \w E^*\op\ot_ZTX\ni u^\la\dr_\la \mapsto u^\la(\dr_\la+\G^i_\la\dr_i
+\g_\la{}^a{}_bc^b\frac{\dr}{\dr c^a})\in \cV_E 
\ee
over $Z$, called a composite graded connection on $Z\to X$.
It is represented by a section
\beq
\g_S= \G + \g_\la{}^a{}_bc^bdx^\la\ot\frac{\dr}{\dr c^a} \label{cmp122}
\eeq
of the fibre bundle $T^*X\op\ot_Z\cV_E\to Z$ such that the composition
\be
Z\op\to^{\g_S}T^*X\op\ot_Z\cV_E\to T^*X\op\ot_Z (\w
E^*\op\ot_Z TZ)\to T^*X\op\ot_ZTX
\ee
is the pull-back onto $Z$ of the canonical form $dx^\la\ot\dr_\la$ on $X$.

Given a graded manifold $(Z,\cA_E)$, the dual of the sheaf $\der\cA_E$ is
the sheaf $\der^*\cA_E$ generated by the $\cA_E$-module morphisms
\beq
\f:\der(\cA_E(U))\to \cA_E(U). \label{z789}
\eeq
One can think of its sections as being graded exterior 1-forms on the graded
manifold $(Z,\cA)$. They are represented by sections of the vector
bundle $\cV^*_E\to 
Z$ which is the $\w E^*$-dual of $\cV_E$. This vector bundle 
is locally isomorphic to the vector bundle
\be
\cV^*_E\mid_U\approx \w E^*\op\ot_Z(\pr_2VE^*\op\oplus_Z T^*Z)\mid_U,
\ee
and has the transition functions
\be
&& v'_{j_1\ldots j_kj}= \rho^{-1}{}_{j_1}^{a_1}\cdots
\rho^{-1}{}_{j_k}^{a_k} \rho^{-1}{}_j^a v_{a_1\ldots a_ka}, \nonumber\\
&& z'_{i_1\ldots i_kA}=
\rho^{-1}{}_{i_1}^{b_1}\cdots
\rho^{-1}{}_{i_k}^{b_k}\left[z_{b_1\ldots b_kA}+ \frac{k!}{(k-1)!} 
v_{b_1\ldots b_{k-1}j}\dr_A\rho^j_{b_k}\right] 
\ee
of the bundle coordinates $(z_{a_1\ldots a_kA},v_{b_1\ldots b_kj})$,
$k=0,\ldots,m$, with respect to the dual bases $\{dz^A\}$ for $T^*Z$ and
$\{dc^b\}$ for $\pr_2V^*E=E^*$. 
Graded exterior 1-forms read 
\be
\f=\f_A dz^A + \f_adc^a.
\ee
They have the coordinate transformation law
\be
\f'_a=\rho^{-1}{}_a^b\f_b, \qquad \f'_A=\f_A
+\rho^{-1}{}_a^b\dr_A(\rho^a_j)\f_bc^j.
\ee
Then
the morphism (\ref{z789}) can be seen as the interior product 
\beq
u\rfloor \f=u^A\f_A + (-1)^{\nw{\f_a}}u^a\f_a. \label{cmp65}
\eeq

There is the exact sequence
\be
0\to \w E^*\op\ot_ZT^*Z\to\cV^*_E\to \w E^*\op\ot_Z \pr_2VE^*\to 0
\ee
of vector bundles.
Any graded connection $\wt\g$ (\ref{cmp93}) yields the
splitting of this exact sequence, and defines the corresponding
decomposition of graded 1-forms
\be
\f=\f_A dz^A + \f_adc^a =(\f_A+\f_a\wt\g_A^a)dz^A +\f_a(dc^a
-\wt\g_A^adz^A). 
\ee

Graded $k$-forms $\f$ are defined as sections
of the graded exterior bundle $\op\w^k_Z\cV^*_E$ such that
\be
\f\w\si =(-1)^{\nm\f\nm\si +\nw\f\nw\si}\si\w \f,
\ee
where $|.|$ denotes the form degree.
The interior product (\ref{cmp65})
is extended to higher degree graded forms by the rule  
\be
u\rfloor (\f\w\si)=(u\rfloor \f)\w \si
+(-1)^{\nm\f+\nw\f\nw{u}}\f\w(u\rfloor\si). 
\ee
The graded exterior differential
$d$ of graded functions is introduced in accordance with the condition 
$u\rfloor df=u(f)$
for an arbitrary graded vector field $u$, and  is
extended uniquely to higher degree graded forms by the rules
\be
d(\f\w\si)= (d\f)\w\si +(-1)^{\nm\f}\f\w(d\si), \qquad  d\circ d=0.
\ee
It takes the coordinate form
\be
d\f= dz^A \w \dr_A(\f) +dc^a\w \frac{\dr}{\dr c^a}(\f), 
\ee
where the left derivatives 
$\dr_A$, $\dr/\dr c^a$ act on the coefficients of graded exterior forms
by the rule 
(\ref{cmp50'}), and they are graded commutative with the forms $dz^A$, $dc^a$.
The Lie
derivative of a graded exterior form $\f$ along a graded vector field
$u$ is given by 
the familiar formula
\be
\bL_u\f= u\rfloor d\f + d(u\rfloor\f). 
\ee

Let $E\to Z$ and $E'\to Z'$ be vector bundles and
$\Phi:E\to E'$
a linear bundle morphism over a morphism $\zeta: Z\to Z'$. 
Then every section $s^*$ of the dual bundle $E'^*\to Z'$ defines the
pull-back section $\Phi^*s^*$ of the dual bundle $E^*\to Z$ by the law
\be
v_z\rfloor\Phi^*s^*(z)= \Phi(v_z)\rfloor s^*(\zeta(z)), \qquad \forall
v_z\in E_z.
\ee
It follows that a linear bundle morphism $\Phi$ yields a morphism 
\beq
S\Phi:(Z,\cA_E)\to (Z',\cA_{E'}) \label{ijmp55}
\eeq
of simple
graded manifolds seen as locally ringed spaces
\cite{bart}. This is the pair $(\zeta, \zeta_*\circ\Phi^*)$
of the morphism $\zeta$ of the
body manifolds and the composition of the pull-back 
$\cA_{E'}\ni f\mapsto \Phi^*f\in \cA_E$ of
graded
functions  and the direct image
$\zeta_*$ of the sheaf $\cA_E$ onto $Z'$. With respect to
local bases $(z^A,c^a)$ and $(z'^A,c'^a)$ for $(Z,\cA_E)$ and
$(Z',\cA_{E'})$, the morphism (\ref{ijmp55}) reads
\be
S\Phi(z)=\zeta(z), \qquad S\Phi(c'^a)=\Phi^a_b(z)c^b.
\ee
Accordingly, the pull-back onto $Z$ of graded exterior forms on $Z'$ is
defined.

\section{SUSY-extended Lagrangian formalism}

The SUSY-extended field theory is constructed as the BRS-generalization of
the vertical extension of field theory on the fibre bundle $VY\to X$ in
Section 4.  

Let us consider the vertical tangent bundle $VVY\to VY$ of $VY\to X$
and the simple graded manifold $(VY,\cA_{VVY})$ whose body manifold 
is $VY$ and the characteristic vector
bundle is $VVY\to VY$. Its local basis is $(x^\la,y^i,\dot y^i,
c^i,\ol c^i)$ 
where $\{c^i,\ol c^i\}$ is the fibre basis for $V^*VY$, dual of the
holonomic fibre basis $\{\dr_i,\dot\dr_i\}$ for $VVY\to VY$. 
Graded vector fields and graded exterior 1-forms are introduced on $VY$
as sections of the vector bundles $\cV_{VVY}$ and $\cV^*_{VVY}$, respectively.
Let us complexify these bundles as $\bC\op\ot_X\cV_{VVY}$ and
$\bC\op\ot_X\cV^*_{VVY}$.
By the BRS
operator on graded functions on $VY$ is meant the complex graded
vector field $u_Q$ (\ref{ijmp30}). It satisfies the nilpotency rule $u_Q^2=0$.

The configuration space of the SUSY-extended field theory is the simple
graded
manifold $(VJ^1Y, \cA_{VVJ^1Y})$ whose characteristic vector bundle is
the vertical tangent bundle $VVJ^1Y\to VJ^1Y$ of $VJ^1Y\to X$. 
Its local basis is $(x^\la,y^i, y^i_\la, c^i,\ol c^i, c^i_\la, \ol c^i_\la)$,
where $\{c^i,\ol c^i, c^i_\la, \ol c^i_\la\}$ is the fibre basis for
$V^*VJ^1Y$ dual of the holonomic fibre basis
$\{\dr_i,\dot\dr_i,\dr^\la_i,\dot\dr^\la_y\}$ for $VVJ^1Y\to VJ^1Y$.
The affine fibration $\pi^1_0:VJ^1Y\to VY$ and the corresponding
vertical tangent morphism $V\pi^1_0:VVJ^1Y\to VVY$ 
yields the
associated morphism of graded manifolds $(VJ^1Y, \cA_{VVJ^1Y})\to
(VY,\cA_{VVY})$ (\ref{ijmp55}). 

Let us introduce the operator of the total derivative
\be
d_\la=\dr_\la +y^i_\la\dr_i + \dot y^i_\la\dot\dr_i
+c^i_\la\frac{\dr}{\dr c^i} + \ol c^i_\la\frac{\dr}{\dr \ol c^i}.
\ee
With this operator, the coordinate transformation laws of $c^i_\la$ and $\ol
c^i_\la$ read
\beq
c'^i_\la= d_\la c'^i, \qquad \ol c'^i_\la= d_\la \ol c'^i. \label{ijmp40}
\eeq
Then one can treat $c^i_\la$ and $\ol c^i_\la$ as the jets of 
$c^i$ and $\ol c^i$. Note that this is not the notion of
jets of graded bundles in \cite{rup}. The transformation laws
(\ref{ijmp40}) show that the BRS operator $u_Q$ (\ref{ijmp30}) on
$VY$ can give rise to the complex graded vector field 
\beq
Ju_Q= u_Q+ c^i_\la\dr_i^\la + i\dot y^i_\la\frac{\dr}{\dr\ol c^i_\la}
\label{ijmp41} 
\eeq
on the $VJ^1Y$. It satisfies the nilpotency rule $(Ju_Q)^2=0$.

In a similar way, the simple graded manifold with the characteristic
vector bundle $VVJ^kY\to VJ^kY$ can be defined. Its local basis is the
collection 
\be
(x^\la,y^i, y^i_\La, c^i,\ol c^i, c^i_\La, \ol c^i_\La), \qquad 0<
|\La|\leq k.
\ee
Let us introduce the operators
\be
&&\dr_c= c^i\dr_i + c^i_\la\dr_i^\la+c^i_{\la\m}\dr_i^\m+\cdots, \qquad 
\dr_{\ol c}= \ol c^i\dr_i + \ol c^i_\la\dr_i^\la +\ol
c^i_{\la\m}\dr_i^\m+\cdots,\\ 
&& d_\la=\dr_\la +y^i_\la\dr_i +c^i_\la\frac{\dr}{\dr c^i} +
\ol c^i_\la\frac{\dr}{\dr \ol c^i} +\cdots.
\ee
It is easily verified that 
\beq
d_\la\dr_c=\dr_cd_\la, \qquad d_\la\dr_{\ol c}=\dr_{\ol c}d_\la.
\label{ijmp44} 
\eeq

As in the BRS mechanics [1-4], the main criterion of the SUSY extension
of Lagrangian formalism is its invariance under the BRS transformation
(\ref{ijmp41}). The BRS-invariant extension of the vertical Lagrangian
$L_V$ (\ref{m18}) is the graded $n$-form
\beq
L_S=L_V + i\dr_c\dr_{\ol c}\cL\om \label{ijmp42}
\eeq
such that $\bL_{Ju_Q}L_S=0$. The corresponding
Euler--Lagrange equations are defined as the kernel of the
Euler--Lagrange operator
\be
\cE_{L_S}= (dy^i \dl_i +d\dot y^i \dot\dl_i + dc^i\frac{\dl}{\dl c^i}
+d\ol c^i\frac{\dl}{\dl \ol c^i})\cL_S\w\om. 
\ee
They read
\bea
&& \dot\dl_i \cL_S=\dl_i\cL=0, \label{ijmp45a}\\
&& \dl_i\cL_S=\dl_i\cL_V +i\dr_{\ol c} \dr_c\dl_i\cL=0, \label{ijmp45b} \\
&& \frac{\dl}{\dl c^i} \cL_S=-i\dr_{\ol c}\dl_i\cL=0, \label{ijmp45c}\\
&& \frac{\dl}{\dl\ol c^i} \cL_S=i\dr_c\dl_i\cL=0, \label{ijmp45d}
\eea
where the relations (\ref{ijmp44}) are used. The equations
(\ref{ijmp45a}) are the Euler--Lagrange equations for the initial
Lagrangian $L$, while  (\ref{ijmp45b}) - (\ref{ijmp45d}) can be seen as
the equations
for a Jacobi field $\dl y^i=\ol\varepsilon c^i +\ol c^i\varepsilon +
i\ol\varepsilon\varepsilon \dot y^i$ modulo terms of order $>2$ in the
odd parameters $\varepsilon$ and $\ol\varepsilon$.

\section{SUSY-extended Hamiltonian formalism}

A momentum phase space of the SUSY-extended field theory is the
complexified simple
graded manifold $(V\Pi,\cA_{VV\Pi})$ whose characteristic vector bundle
is $VV\Pi \to V\Pi$ \cite{nuo}. 
Its local basis is
\be
(x^\la,y^i,p^\la_i,\dot y^i,\dot p^\la_i, c^i, \ol c^i, c^\la_i,\ol c^\la_i),
\ee
where
$c_i^\la$  and $\ol c_i^\la$ have
the same transformation laws as $p_i^\la$ and $\dot p_i^\la$, respectively.
The corresponding 
graded vector fields and graded 1-forms are introduced on $V\Pi$ as
sections of the 
vector bundles $\bC\op\ot_X\cV_{VV\Pi}$ and
$\bC\op\ot_X\cV^*_{VV\Pi}$, respectively.  

In accordance with the above mentioned transformation laws of 
$c_i^\la$  and $\ol c_i^\la$, 
the BRS operator $u_Q$ (\ref{ijmp30}) on
$VY$ can give rise to the complex graded vector field 
\beq
\wt u_Q= \dr_c +i\dot y^i
\frac{\dr}{\dr \ol c^i} + i\dot p_i^\la\frac{\dr}{\dr\ol c_i^\la}
\label{ijmp50} 
\eeq
on $V\Pi$. The BRS-invariant extension of the polysymplectic form  
$\Om_{VY}$ (\ref{cmp35}) on $V\Pi$ is the
$TX$-valued graded form
\be
\Om_S=[d\dot p_i^\la\w dy^i +dp_i^\la\w d\dot y^i +i(d\ol c_i^\la\w dc^i- d\ol
c^i\w dc_i^\la)]\w \om\ot\dr_\la,
\ee
where $(c^i,-i\ol c^\la_i)$ and $(\ol c^i,i c^\la_i)$ are the
conjugate pairs.
Let $\g$ be a Hamiltonian connection for a Hamiltonian form $H$ on $\Pi$. 
Its double
vertical prolongation $VV\g$ on $VV\Pi\to X$ is a linear morphism over the
vertical connection $V\g$ on $V\Pi\to X$, and so defines the composite
graded connection 
\be
(VV\g)_S =V\g + dx^\m\ot[\ol g^i_\m\frac{\dr}{\dr \ol c^i} +\ol
g^\la_{\m i}\frac{\dr}{\dr\ol c_i^\la} + g^i_\m\frac{\dr}{\dr c^i} +g^\la_{\m
i}\frac{\dr}{\dr c_i^\la}]
\ee
(\ref{cmp122}) on $V\Pi\to X$, whose components  $g$ and $\ol g$ are given by
the expressions
\be
&& \ol g^i_\la=\dr_{\ol c}\dr^i_\la\cH, \quad
\ol g^\la_{\la i}=-\dr_{\ol c}\dr_i\cH,\quad
g^i_\la= \dr_c\dr^i_\la\cH,
\quad g^\la_{\la i}= -\dr_c\dr_i\cH, \\
&& \dr_c=c^i\dr_i + c_i^\la\dr^i_\la, \qquad
  \dr_{\ol c}=\ol c^i\dr_i + \ol c_i^\la\dr^i_\la.
\ee
This composite graded connection satisfies the relation
\be
(VV\g)_S\rfloor\Om_S=-dH_S,
\ee
and can be regarded as a Hamiltonian graded connection for the Hamiltonian
graded form 
\ben
&& H_S=[\dot p_i^\la dy^i +p_i^\la d\dot y^i +i(\ol c_i^\la dc^i +d\ol
c^i c_i^\la)]\om_\la -\cH_S\om, \label{z750}\\
&& \cH_S=(\dr_V 
+ i\dr_{\ol c}\dr_c)\cH, \nonumber
\een
on $V\Pi$. It is readily observed that this graded form is
BRS-invariant, i.e., $\bL_{\wt u_Q}
H_S=0$. Thus, it is the desired SUSY extension of the Hamiltonian form $H$.

The Hamiltonian graded form $H_S$ (\ref{z750}) defines the
corresponding SUSY extension of the Lagrangian $L_H$ (\ref{Q3}) as follows. 
The fibration $J^1V\Pi\to V\Pi$ yields the pull-back
of the 
Hamiltonian graded form $H_S$ (\ref{z750}) onto $J^1V\Pi$.
Let us consider the graded generalization of the operator $h_0$
(\ref{cmp100}) such that 
\be
h_0: dc^i\mapsto c^i_\m dx^\m, \quad dc^\la_i\mapsto c^\la_{\m i}dx^\m.
\ee
Then the graded horizontal 
density
\ben
&& L_{SH}=h_0(H_S)= (L_H)_S=L_{H_V} + i[(\ol c_i^\la
c^i_\la+\ol c^i_\la c_i^\la)  -\dr_{\ol c}\dr_c\cH]\om=  L_{H_V}
+\label{cmp104}\\ 
&& \qquad i[\ol c^\la_i(c^i_\la -\dr_c\dr^i_\la\cH) + (\ol
c^i_\la -\dr_{\ol c}\dr^i_\la\cH)c^\la_i + \ol
c^\la_ic^\m_j\dr^i_\la\dr^j_\m\cH -
\ol c^ic^j\dr_i\dr_j\cH]\om
\nonumber
\een
on $J^1V\Pi\to X$ is the SUSY  extension (\ref{ijmp42}) of the
Lagrangian $L_H$ (\ref{Q3}). The Euler--Lagrange equations for $L_{SH}$
coincide with the Hamilton equations for $H_S$, and read
\bea
&& y^i_\la=\dot\dr^i_\la \cH_S=\dr^i_\la\cH, \qquad 
p^\la_{\la i}= -\dot\dr_i\cH_S=-\dr_i\cH, \label{ijmp51a}\\
&& \dot y^i_\la=\dr^i_\la\cH= (\dr_V +i\dr_{\ol c}\dr_c)\dr^i_\la\cH,
\qquad \dot p^\la_{\la i}=-\dr_i\cH_S=-(\dr_V+i\dr_{\ol
c}\dr_c)\dr_i\cH, \label{ijmp51b}\\
&& c^i_\la =i\frac{\dr\cH_S}{\dr \ol c^\la_i}= -\dr_c\dr^i_\la\cH, \qquad
c^\la_{\la i} =i\frac{\dr\cH_S}{\dr \ol c^i}= -\dr_c\dr_i\cH,
\label{ijmp51c}\\ 
&& \ol c^i_\la =-i\frac{\dr\cH_S}{\dr \ol c^\la_i}= -\dr_{\ol
c}\dr^i_\la\cH, \qquad 
\ol c^\la_{\la i} =-i\frac{\dr\cH_S}{\dr \ol c^i}= -\dr_{\ol c}\dr_i\cH.
\label{ijmp51d}
\eea
The equations (\ref{ijmp51a}) are the Hamilton equations for the
initial Hamiltonian form $H$, while (\ref{ijmp51b}) -- (\ref{ijmp51d})
describe the Jacobi fields
\be
\dl y^i=\ol\varepsilon c^i+\ol c^i\varepsilon+i\ol
\varepsilon\varepsilon\dot y^i, \qquad \dl p^\la_i=\ol\varepsilon c^\la_i+\ol
c^\la_i\varepsilon+i\ol \varepsilon\varepsilon\dot p^\la_i.
\ee

Let us study the relationship between SUSY-extended Lagrangian and
Hamiltonian formalisms.
Given a Lagrangian $L$ on $J^1Y$,
the vertical Legendre map $\wh L_V$ (\ref{ijmp56}) yields the
corresponding morphism
(\ref{ijmp55}) of graded manifolds
\be
S\wh L_V: (VJ^1Y, \cA_{VVJ^1Y})\to (V\Pi, \cA_{VV\Pi})
\ee
which is given by the relations (\ref{ijmp57}) and 
\be
c^\la_i=\dr_c\pi^\la_i, \qquad \ol c^\la_i=\dr_{\ol c}\pi^\la_i.
\ee
Let $H$ be a Hamiltonian form on $\Pi$. The vertical
Hamiltonian map $\wh H_V$ (\ref{ijmp58}) yields the morphism of graded
manifolds 
\be
S\wh H_V: (V\Pi, \cA_{VV\Pi}) \to (VJ^1Y, \cA_{VVJ^1Y})
\ee
given by the relations (\ref{ijmp59}) and 
\be
c^i_\la= \dr_c\dr^i_\la\cH, \qquad \ol c^i_\la= \dr_{\ol c}\dr^i_\la\cH.
\ee
If a Hamiltonian form $H$ is associated with $L$, a direct computation
shows that the Hamiltonian graded form $H_S$ (\ref{z750}) is weakly
associated with the Lagrangian $L_S$ (\ref{ijmp42}), i.e.,
\be
&& S\wh L_V\circ S\wh H_V\circ S\wh L_V=S\wh L_V,\\
&& L_S\circ S\wh H_V= (p^\la_i\dr^i_\la +\dot p^\la_i\dot \dr^i_\la +
c^\la_i \frac{\dr}{\dr c^\la_i} + \ol c^\la_i \frac{\dr}{\dr \ol
c^\la_i})\cH_S -\cH_S,
\ee
where the second equality  takes place at points of the Lagrangian
constraint space $\wh L(J^1Y)$.

\section{The BRS-invariance}

Now turn to the above mentioned case of fibre bundles $Y\to X$ with
affine transition functions. Since
transition functions of the holonomic coordinates $\dot y^i$ on $VY$ are 
independent of $y^i$, the transformation laws of the frames $\{\dr_i\}$
and $\{\dot \dr_i\}$ are the same, and so are the transformations laws of
the coframes $\{c^i\}$ and $\{\ol c^i\}$. Then the graded vector fields
(\ref{cmp131}) are globally defined on $VY$. The graded vector fields 
(\ref{ijmp30}) and (\ref{cmp131}) 
constitute the above-mentioned
Lie superalgebra of the  supergroup ISp(2):
\ben
&& [u_Q,u_Q]=[u_{\ol Q},u_{\ol Q}]=[u_{\ol Q}, u_{Q}]
=[u_K,u_Q]=[u_{\ol K},u_{\ol Q}]=0, \nonumber\\
&& [u_K,u_{\ol Q}]=u_Q, \qquad [u_{\ol K},u_Q]=u_{\ol Q}, \quad
[u_K,u_{\ol K}]= u_C, \label{cmp128} \\
&& [u_C,u_K]=2u_K, \quad [u_C,u_{\ol K}]=-2u_{\ol K}. \nonumber
\een

Similarly to (\ref{ijmp41}), let us consider the jet prolongation of the
graded vector fields (\ref{cmp131}) onto $VJ^1Y$. Using the compact
notation $u=u^a\dr_a$, we have the formula
\be
Ju=u +d_\la u^a\dr_a^\la
\ee
and, as a consequence, obtain
\ben
&& Ju_{\ol Q}=u_{\ol Q} + \ol c^i_\la\dr_i^\la  -i\dot y^i_\la
\frac{\dr}{\dr c^i_\la}, \nonumber  \\
&& Ju_K=u_K+ c_\la^i\frac{\dr}{\dr \ol c_\la^i}, \qquad
Ju_{\ol K}=u_{\ol K}+ \ol c_\la^i\frac{\dr}{\dr c_\la^i}, \label{ijmp60}\\
&& Ju_C=u_C =c^i_\la\frac{\dr}{\dr c^i_\la}  -\ol c^i_\la
\frac{\dr}{\dr\ol c^i_\la}. \nonumber
\een
It is readily observed that the SUSY-extended Lagrangian $L_S$
(\ref{ijmp42}) is invariant under the transformations (\ref{ijmp60}).
The graded vector fields (\ref{ijmp41}) and (\ref{ijmp60}) make up the
Lie superalgebra (\ref{cmp128}).

The graded vector fields (\ref{cmp131}) can give rise to
$V\Pi$ by the formula
\be
\wt u=u - (-1)^{[y^a]([p_b] +[u^b])}\dr_au^bp^\la_b\frac{\dr}{\dr p_a^\la}.
\ee
We have
\ben
&& \wt u_{\ol Q}= \dr_{\ol c} -i\dot y^i
\frac{\dr}{\dr c^i} - i\dot p_i^\la\frac{\dr}{\dr c_i^\la}, \nonumber\\
&& \wt u_K= c^i\frac{\dr}{\dr \ol
c^i} +c_i^\la\frac{\dr}{\dr \ol c_i^\la}, \qquad
\wt u_{\ol K}= \ol c^i\frac{\dr}{\dr
c^i} +\ol c_i^\la\frac{\dr}{\dr c_i^\la},
\label{cmp130}\\
&& \wt u_C= c^i\frac{\dr}{\dr c^i} + c_i^\la\frac{\dr}{\dr c_i^\la} - 
 \ol c^i\frac{\dr}{\dr\ol c^i} - \ol c_i^\la\frac{\dr}{\dr \ol c_i^\la}.
\nonumber
\een
A direct computation shows that the BRS-extended Hamiltonian form 
$H_S$ (\ref{z750}) is invariant under the transformations (\ref{cmp130}).
Accordingly, the Lagrangian $L_{SH}$ (\ref{cmp104}) is invariant under
the jet prolongation $J\wt u$ of the graded vector fields (\ref{cmp130}).
The graded vector fields (\ref{ijmp50}) and (\ref{cmp130}) make up the
Lie superalgebra (\ref{cmp128}).

With the graded vector fields (\ref{ijmp50}) and (\ref{cmp130}), one
can construct the corresponding graded currents
$J_u=\wt u\rfloor H_S=u\rfloor H_S$. These are the graded 
$(n-1)$-forms
\be
&& Q=(c^i\dot p^\la_i-\dot y^i c_i^\la)\om_\la, \quad \ol Q=(\ol c^i\dot
p^\la_i-\dot y^i \ol c_i^\la)\om_\la, \\
&& K=-i c_i^\la c^i\om_\la, \quad \ol K=i\ol c_i^\la \ol c^i\om_\la,\quad
C=i(\ol c_i c^i -\ol c^i c_i)\om_\la
\ee
on $V\Pi$. They form the Lie superalgebra (\ref{cmp128}) with respect to the
product
\be
[J_u,J_{u'}]=J_{[u,u']}.
\ee

The following construction is similar to that in the SUSY and
BRS mechanics.
Given a function $F$ on the Legendre bundle $\Pi$,  
let us consider the operators
\be
F_\bt=e^{\bt F}\circ \wt u_Q\circ e^{-\bt F} =\wt u_Q
-\bt\dr_c F, \quad 
\ol F_\bt=e^{-\bt F}\circ \wt u_{\ol Q}\circ e^{\bt F}=\wt u_{\ol Q}
+\bt\dr_{\ol c} F, \quad \bt>0, 
\ee
called the SUSY charges,  which act on graded functions
on $V\Pi$. These operators are nilpotent, i.e.,
\beq
F_\bt\circ F_\bt=0, \qquad \ol F_\bt\circ \ol F_\bt=0. \label{z763}
\eeq
By the BRS-invariant extension of a function $F$ is meant the
graded function
\be
 F_S= -\frac{i}{\bt}(\ol F_\bt\circ F_\bt +F_\bt\circ \ol F_\bt).
\ee
We have the relations 
\be
F_\bt\circ F_S- F_S\circ F_\bt=0, \qquad
\ol F_\bt\circ F_S- F_S\circ\ol F_\bt=0. 
\ee
These relations together with the relations (\ref{z763}) provide the
operators
$F_\bt$, $\ol F_\bt$, and $F_S$ with the structure of the Lie
superalgebra sl(1/1) \cite{cec}.
In particular, let $F$ be a local function $\cH$ in the expression
(\ref{b418}). Then 
\be
\cH_S=-i(\ol F_1\circ F_1 +F_1\circ \ol F_1) 
\ee
is exactly the local function $H_S$ in the expression (\ref{z750}).
The similar splitting of a
super-Hamiltonian is the corner stone  of the SUSY  mechanics
\cite{lah,coop}.

\end{document}